\documentclass[12pt]{article}
\usepackage{graphicx}
\usepackage[top = 2cm]{geometry}
\usepackage{hyperref}

\begin{document}

\def\onehalf{{\textstyle \frac12}}
\def\ii{{\rm i}}
\def\dd{{\rm d}}
\def\jour#1#2#3#4{{\it #1{}} {\bf #2}, #3 (#4)}
\def\lab#1{\label{eq:#1}}
\def\rf#1{(\ref{eq:#1})}
\def\Lie#1{\hbox{\sf #1}}
\def\FFF{{\sf F}}
\def\rectangulo#1{\centerline{\framebox{\LARGE #1}}}
\def\of#1{{\scriptstyle(}#1{\scriptstyle)}}
\def\oof#1{{\scriptscriptstyle(}#1{\scriptscriptstyle)}}
\def\tsty#1#2{{\textstyle\frac{#1}{#2}}}
\def\ssty#1{{{\scriptscriptstyle #1}}}
\def\abs#1{{\scriptstyle|}#1{\scriptstyle|}}
\def\vecdos#1#2{\bigg({#1 \atop #2}\bigg)}
\def\matdos#1#2#3#4{\bigg({#1 \atop #3}\ {#2\atop #4}\bigg)}
\def\mattres#1#2#3#4#5#6#7#8#9{\left(\begin{array}{ccc}
	#1&#2&#3\\ #4&#5&#6\\ #7&#8&#9 \end{array} \right)}	
\def\matricita#1#2#3#4{{\textstyle \Big({#1 \atop #3}\ {#2\atop #4}\Big)}}
\def\Re{{\rm Re}\,}  \def\Im{{\rm Im}\,}
\def\ket#1#2{| #1 \rangle_{\scriptscriptstyle #2}}
\def\bra#1#2{{}_{\scriptscriptstyle #1}\langle #2 |}
\def\braket#1#2#3#4{{}_{\scriptscriptstyle #1}\langle #2 | #3 \rangle_{\scriptscriptstyle #4}}
\def\ssr#1{{\scriptscriptstyle\rm #1}}

\newcommand{\be}{\begin{equation}}
\newcommand{\ee}{\end{equation}}
\newcommand{\bea}{\begin{eqnarray}}
\newcommand{\eea}{\end{eqnarray}}

\begin{center}
{\LARGE Unitary rotation of\\ pixellated polychromatic images}\\[20pt] 	
 {Alejandro R.\ Urz\'ua} and {Kurt Bernardo Wolf}\\[15pt]
{Instituto de Ciencias F\'{\i}sicas\\
Universidad Nacional Aut\'onoma de M\'exico\\Av.\ Universidad s/n, 
Cuernavaca, Morelos 62251, M\'exico}
\end{center} 

\vskip25pt

\begin{abstract}
Unitary rotations of polychromatic images on finite 
two-dimen\-sional pixellated screens provide invertibility, 
group composition, and thus conservation of information. 
Rotations have been applied on monochromatic image data 
sets, where we now examine closer the Gibbs-{\it like\/} 
oscillations that appear due to discrete `discontinuities'
of the input images under unitary transformations. Extended
to three-color images we examine here the display of color 
at the pixels where, due to the oscillations, some pixel color
values may fall outside their required common numerical range 
$[0,1]$, between absence and saturation of the red, green, 
and blue formant color images. 
\end{abstract}


\section{Introduction}       \label{sec:one}  

In geometric and in wave optics, transformations that respect the
Hamiltonian structure of the models are called canonical; they are 
termed linear when the phase space coordinates are linearly mapped. 
Images displayed on screens can undergo linear maps representing 
free propagation, rotation, gyration, squeezing, 
fractional Fourier and other transformations for one-  
or two-dimensional images on line or plane screens.

The {\it discrete\/} version of these systems and transformations, are based on the 
finite oscillator model \cite{AW,APVW-II,VW-rot}, which considers finite {\it pixellated\/} 
line or plane screens that bear the set of generally complex data values that form 
the image.  The requirement that the transformations be {\it unitary\/} to conserve 
Hilbert space properties of the image data sets (orthogonality, reality), requires 
them to be elements of a {\it group}, of which screen rotations form a one-parameter 
subgroup, to guarantee invertibility and proper concatenation for two or more such 
transformations. In particular, when the pixel values are real, they should remain 
so under rotation. 
 
For this article to be essentially self-contained, in Sect.\ \ref{sec:two} 
we remind the reader of the basic finite oscillator 
model for $D=1$ and $D=2$-dimensional finite pixellated images along Cartesian 
coordinates \cite{APVW-I}. We emphasize that this rotation 
algorithm provides the {\it only\/} unitary rotation coefficients for the 
image pixels on an $N_x\times N_y$ generally rectangular screen \cite{rectangular}. 
For simplicity we address here only {\it square\/} $N\times N$ images and screens.
 
In this paper we extend unitary rotations to {\it polychromatic\/} pixellated arrays that
carry the three component color values (for red, green and blue) as generally used in 
the applied literature.  In the polychromatic case, the ranges of the pixel values 
are restricted to the {\it common\/} interval $[0,1]$, between absence and saturation 
of each color. The main problem that appears now is that of Gibbs-{\it like\/} 
oscillations in one or more color component images of pixel values 
that may fall {\it outside\/} that basic $[0,1]$ range. This must be addressed with an
appropriate display tactic, which is laid out in Sect.\ \ref{sec:three}.
Finally, in the concluding Sect.\ \ref{sec:four}, we comment further on 
linear unitary transformations of pixellated data that are required to conserve information
and Hamiltonian structure.


\section{Finite one- and two-dimensional images}    \label{sec:two}

The finite oscillator model \cite{finite-model} is based 
on the Lie algebra of spin \Lie{su($2$)}, with generators $\{ J_1,J_2,J_3\}$
of commutation relations $[J_i,J_j]=\varepsilon_{i,j,k}J_k$ ($i,j,k$ cyclic), 
whose representation multiplets, of dimension $N=2j+1$ (with $j$ 
a non-negative integer or half-integer), are interpreted to contain its
$N$ equidistant eigenstates. The eigenvalues of the 
generator $Q:=J_1$ characterize the $N$ pixel {\it positions\/} on 
a screen, $P:=J_2$ is momentum, while those of $K:=J_3$ number the 
$N$ `energy' states.
The overlap between the position and mode eigenbases yields 
the finite oscillator {\it wavefunctions}. This finite model mimics
the usual `continuous' harmonic oscillator based on the 
Heisenberg-Weyl algebra (of commutator 
$[\overline{Q},\overline{P}]=\ii\hbar{\it 1}$, where the 
spectrum of position  $\overline{Q}$ and of momentum 
$\overline{P}$ are the full real line), whose energy eigenstates of 
$\overline{H}:=\onehalf(\overline{P}^2+\overline{Q}^2)$ 
are equally-spaced with only a lower bound.

\subsection{The one-dimensional finite oscillator and screen}

For  $D=1$-dimensional screens of $N$ pixels in a line, the 
{\it finite quantization\/} of the harmonic oscillator replaces 
the commutator $[\overline{Q},\overline{P}]=\ii{\it1}$ 
of the continuous model with $[Q,P]=\ii K$ in their 
$N=2j+1$-dimensional representation \cite{AW}. 
Finite quantization thus assigns $N\times N$ \Lie{su($2$)} 
matrix representations $\bf Q$ and $\bf P$ to position and
momentum $q$ and $p$, as follows:
\bea
	&&{\hskip-15pt}\hbox{position}:\quad q\mapsto Q := J_1,  \quad{\bf Q}=\Vert Q_{q,q'}\Vert, \lab{defQ}\\
	&&Q_{q,q'} = q\,\delta_{q,q'}, \qquad q,q' \in\{-j,-j{+}1,\ldots\, ,j\},\quad N=:2j+1,\nonumber \\
	&&{\hskip-15pt}\hbox{momentum}:\quad  p\mapsto P := J_2,  \quad{\bf P}=\Vert P_{q,q'}\Vert, \lab{defP}\\
	&&P_{q,q'} = -\ii\onehalf\sqrt{(j{-}q)(j{+}q{+}1)}\,\delta_{q+1,q'}
	           +\ii\onehalf\sqrt{(j{+}q)(j{-}q{+}1)}\,\delta_{q-1,q'},   \nonumber\\
    &&{\hskip-15pt}\hbox{and their commutator}:\quad  K := J_3, 
    						\quad{\bf K}=\Vert K_{q,q'}\Vert,\lab{defK}\\
	&&K_{q,q'} = \onehalf\sqrt{(j{-}q)(j{+}q{+}1)}\,\delta_{q+1,q'}
	           +\onehalf\sqrt{(j{+}q)(j{-}q{+}1)}\,\delta_{q-1,q'}.  \nonumber
\eea
These three operators are thus represented by self-adjoint $N\times N$ 
matrices in the vector space with inner product $\sum_{q=-j}^j f_q g_q^*$,
and  generate the Lie algebra \Lie{su($2$)} with commutation relations 
\be 
	[J_i,J_j]= \ii\,\varepsilon_{i,j,k}J_k, \quad \hbox{i.e.},\quad
	[K,Q]=-\ii P, \ [K,P]=\ii Q, \ [Q,P]=\ii K, \lab{comm-rel}
\ee
where the first two will stand in place of the two Hamiltonian commutators
in the continuous model, $[\overline{H},\overline{Q}]=-\ii\overline{P}$ 
and $[\overline{H},\overline{P}]=\ii \overline{Q}$.
The eigenvalue spectra of the three \Lie{su($2$)} generators  
are $\Sigma(Q)=\Sigma(P)=\Sigma(K)=\{-j,-j{+}1,\ldots\, ,j\}=: |_{-j}^j$,
and we remark that the position operator $Q=J_1$, is here {\it diagonal}.
Finally, the eigenvalues of the {\it mode\/} operator $K+j{\it1}$
will count its eigenstates by the {\it mode number\/} $n:=\kappa+j$ as
\be \
	 \Sigma(K+j{\it1}):=\kappa+j = n\in\{0,\,1,\,\ldots,\,2j\}.
					\lab{moden}
\ee

Using Dirac notation, let the orthogonal and complete eigenbases of the 
position and mode operators, $J_1$ and $J_3$, be
\be 
	Q\ket{j,q}1 = q\ket{j,q}1,\ q|_{-j}^j, \qquad 
		K\ket{j,n}3 = \kappa\ket{j,n}3 =(n{-}j)\ket{j,n}3,\ n|_0^{2j}.	\lab{ket-qn}
\ee
One-dimensional finite signals or discrete {\it images\/} ${\bf F}=\{F_q\}_{q=-j}^j$ 
are then given by $N=(2j{+}1)$-dimensional vectors of components
\be 
	F_{-j}=\braket1{j,-j}{{\bf F}}{},\quad F_{-j+1}=\braket1{j,-j{+}1}{{\bf F}}{},
		\quad\cdots\quad F_{j}=\braket1{j,j}{{\bf F}}{}.  \lab{Fimage1}
\ee	
The overlap between the two bases in \rf{ket-qn} leads to a three-term difference 
equation that yields the Wigner
`little-$d$' functions \cite{Bied-Louck} for the angle $\onehalf\pi$ 
between the two axes, which are the {\it finite oscillator\/} wavefunctions 
\cite{AW}: 
\be
	 \begin{array}{rcl}
	\Psi_n(q) &:=& \braket1{j,q}{j,n}3 = d^j_{n-j,q}(\onehalf\pi) 
			=\Psi_{q+j}(n{-}j) \\[7pt]
	&=& \displaystyle\frac{(-1)^n}{2^j} 
		\sqrt{ \bigg({2j\atop n}\bigg)\bigg({2j \atop q{+}j}\bigg)}
			\, K_n(q{+}j;\onehalf;2j), \end{array}\lab{Kravch} 
\ee
that are given in terms of the square root of a binomial coefficient
in $q$, a discrete version of a Gaussian, and symmetric Kravchuk
polynomials of degree $n$, 
$K_n(q{+}j;\onehalf;2j)=K_{q+j}(n;\onehalf;2j)={}_2F_{\!1}(-n,-2j{-}q;-2j;2)$,
that are discrete analogues of Hermite polynomials. The set of 
functions $\{\Psi_n(q)\}_{n=0}^{2j}$, and also the set $\{\Psi_n(q)\}_{q=-j}^{j}$, 
form orthonormal bases for the $N$-dimensional space of signals (i.e., 
one-dimensional pixellated images on linear screens), so 
$\sum_{q=-j}^j\Psi_n(q)\Psi_{n'}(q)= \delta_{n,n'}$ and 
$\sum_{n=0}^{2j}\Psi_n(q)\Psi_n(q')= \delta_{q,q'}$.

Finally, we write the general form of the Wigner `little-$d$' 
functions \cite[Sect.\ 3.6]{Bied-Louck},
\be 
	\begin{array}{rcl}
		d^j_{m',m}(\beta)&:=& d^j_{m,m'}(-\beta) 
			= \sqrt{(j+m')(j-m')(j+m)(j-m)} \\[5pt]
		&\times& \displaystyle\sum_k \frac{(-1)^{m'-m+k} 
			(\cos\onehalf\beta)^{2j+m-m'-2k} (\sin\onehalf\beta)^{m'-m+2k}
			}{k!\,(j+m-k)!\,(m'-m+k)!\,(j-m'-k)! },
				\end{array} \lab{Wds}
\ee 
where the summation extends over the integer range
$\hbox{max}(0,m{-}m')\le k \le \hbox{min}(j{-}m)$ 
for $m>m'$, while for $m<m'$ the reflection formula 
in \rf{Kravch} applies.


\subsection{Finite two-dimensional images}    

In $D=2$ dimensions, the finite oscillator Lie algebra is 
$\Lie{su($2$)}_x \oplus\; \Lie{su($2$)}_y=\Lie{so($4$)}$,
i.e., the four-dimensional orthogonal Lie algebra, which 
contains in particular the generator of an \Lie{SO($2$)} 
group of rotations in the $x$-$y$ plane. (Lie algebras are denoted by 
lower-case letters, their generated Lie groups by upper-case letters.)
On a square screen, the \Lie{so($4$)} algebra represents
finite images $\bf F$ by $N\times N$ matrices of entries 
$\{ F_{q_x,q_y}\}$, where $-j\le q_x, q_y \le j$ are integer-spaced.
Generic rectangular pixellated screens $N_x\times N_y$ have 
been considered in \cite{rectangular} with general representations
of \Lie{so($4$)}, but square ones are simpler and suffice for 
this study. 

Finite 2D images can be characterized 
also by a {\it second\/} set of $N^2$ pixel values 
that follow polar coordinates responding to the subalgebra chain 
$\Lie{so($4$)}\supset \Lie{so($3$)} \supset \Lie{so($2$)}$
\cite{APVW-II}. The rotation of pixellated images in a square 
screen is then accomplished by transforming from the Cartesian 
to the polar basis, rotating through multiplicative phases in the
polar basis, and transforming back to the Cartesian basis. 

In two dimensions marked by orthogonal axes $x,\,y$, the
algebra $\Lie{su($2$)}_x \oplus\; \Lie{su($2$)}_y$ provides
two sets of generators \rf{defQ}--\rf{defK} that mutually commute
and provide, as in \rf{ket-qn}, their common eigenbases with $j_x=j=j_y$
(where we omit writing $j$ henceforth),
\be 
	\begin{array}{ll}
	Q_x\ket{q_x,q_y}1 = q_x\ket{q_x,q_y}1,\ q_x|_{-j}^j, \quad &
		K_x\ket{n_x,n_y}3 =(n_x{-}j)\ket{n_x,n_y}3,\ n_x|_{0}^{2j},\\
	Q_y\ket{q_x,q_y}1 = q_y\ket{q_x,q_y}1,\ q_y|_{-j}^j, \quad &
		K_y\ket{n_x,n_y}3 =(n_y{-}j)\ket{n_x,n_y}3,\ n_y|_{0}^{2j},
		  \end{array}  \lab{xxyy}
\ee
defining the $N^2$ states $\{\ket{q_x,q_y}1\}$ of the 
position basis and the $N^2$ states $\{\ket{n_x,n_y}3\}$ 
of the mode basis. The overlap between these two orthogonal 
eigenbases  provides the 2D finite {\it Cartesian\/} oscillator 
wavefunctions,
\be 
	\Psi_{n_x,n_y}(q_x,q_y) =\braket1{q_x,q_y}{n_x,n_y}3 
	= \Psi_{n_x}(q_x)\,\Psi_{n_y}(q_y).  \lab{cart-bas}
\ee
As before, the set of functions $\{\Psi_{n_x,n_y}(q_x,q_y)\}$ forms
orthonormal bases for the $N^2$-dimensional space of pixellated
two-dimensional images with respect to the indices $q_x,q_y$ and 
with respect to $n_x,n_y$.

Two-dimensional images ${\bf F}=\{F_{q_x,q_y}\}_{q_x,q_y=-j}^j$ are thus 
represented by the  values of  
$N\times N$ matrices built with the adjoint of the position basis $\ket{q_x,q_y}{1}$ 
in \rf{xxyy}, as was done in \rf{Fimage1} for one dimension, 
\be 
	{\bf F}= \mattres{\bra{1}{-j,-j}}{\cdots}{\bra{1}{-j,j}
		}{\vdots}{\ddots}{\vdots
		}{\phantom{-}\bra{1}{j,-j}}{\cdots}{\phantom{-}\bra{1}{j,j}}\!\ket{{\bf F}}{}
	=\mattres{F_{-j,-j}}{\cdots}{F_{-j,j}
		}{\vdots}{\ddots}{\vdots
		}{F_{j,-j}}{\cdots}{F_{j,j}},
\ee

Next, according to the commutation relations \rf{comm-rel}, $K_x$
and $K_y$ generate independent rotations in the $(Q_x,P_x)$ and
$(Q_y,P_y)$ phase planes. Their sum and difference have the eigenvalue sets
\bea
	K:=K_x + K_y, && \quad n:=n_x+n_y\in\{0,1,\ldots,4j\}, \lab{specK}\\ 
	M:=K_x - K_y, && \quad\! m:=n_x-n_y\in\{-2j,-2j{+}1,\ldots,2j\}.\lab{specM}
\eea
There follows the definition of a {\it distinct\/} position-momentum 
basis \cite{APVW-II} that is classified by the \Lie{so($4$)} generators
\be 
	Q^\circ_x:=Q_x+Q_y,\quad Q^\circ_y:=P_x-P_y, \qquad
	P^\circ_x:=P_x+P_y, \quad P^\circ_y:=-Q_x+Q_y. \lab{qp-circ}
\ee	
With $K$ and $M$, these operators satisfy the commutation relations 
\be \begin{array}{l}
	\Big[ K, \,\Big\{ {Q_x^\circ\atop P_x^\circ} \Big] = \ii\Big\{ {\phantom{-}P_x^\circ\atop -Q_x^\circ,}\quad
	\Big[ K, \Big\{ {Q_y^\circ\atop P_y^\circ} \Big] = \ii\Big\{ {\phantom{-}P_y^\circ\atop -Q_y^\circ,}\\[5pt]
	\Big[ M, \Big\{ {Q_x^\circ\atop Q_y^\circ} \Big] = \ii\Big\{ {\phantom{-}Q_y^\circ\atop -Q_x^\circ,}\quad
	\Big[ M, \Big\{ {P_x^\circ\atop P_y^\circ} \Big] = \ii\Big\{ {\phantom{-}P_y^\circ\atop -P_x^\circ,}
		\end{array}  	\qquad    [K,M]=0, \lab{pol-bas}
\ee
that interpret  $K$  as the generator of 
isotropic fractional Fourier-Kravchuk transforms 
\cite{APVW-II,VW-rot}, and  $M$ as the generator of {\it rotations\/}
of the phase space operators in \rf{qp-circ}. The commuting operators 
$K,\,M$ define the orthonormal {\it polar\/} eigenbasis $\{\ket{n,m}P\}$,
\be 
	K\ket{n,m}P = (n-2j)\ket{n,m}P,\ n|_0^{4j}, \quad 	
	M\ket{n,m}P =m\ket{n,m}P,\ m|_{-2j}^{2j}.
		\lab{polar-basis}
\ee

The overlap between the position basis and the polar basis determines
the  2D finite {\it polar\/} oscillator wavefunctions \cite{APVW-II},
\be 
	\begin{array}{rcl}
		&&\Lambda^j_{n,m}(q_x,q_y):=\braket1{q_x,q_y}{n,m}P \\[5pt]
		&&{\qquad}=\displaystyle (-1)^{(|m|-m)/2}\! \!\!\sum_{n_x+n_y=n}\!\!\!
		(-\ii)^{n_y}  d^{n/2}_{\frac12(n_x-n_y),\frac12 m}(\onehalf\pi)\,
		\Psi_{n_x,n_y}(q_x,q_y).  \end{array} \lab{radial-mode}
\ee
Here also, the set of functions $\{\Lambda^j_{n,m}(q_x,q_y)\}$ provide
orthonormal bases for the $N^2$-space of images, with respect to
the indices $q_x,q_y$ and with respect to $n,m$.

By construction, the functions \rf{radial-mode} transform 
under rotations through multiplication by phases,  
\be 
	{\cal R}(\theta)\Lambda^j_{n,m}(q_x,q_y) 
		:= e^{-\ii\theta M}\Lambda^j_{n,m}(q_x,q_y)
		= e^{-\ii m\theta}\Lambda^j_{n,m}(q_x,q_y),
		\lab{rot-trans}
\ee
The rotation of finite images ${\cal R}(\theta):\ket{{\bf F}}{}\mapsto
\ket{{\bf F}^{(\theta)}}{}$ from the set 
$\{F_{q_x,q_y}\}$ to $\{F_{\!q_x,q_y}^{(\theta)}\}$ can thus proceed as
\bea
	F_{\!q_x,q_y}^{(\theta)}
		&=&  \bra{1}{q_x,q_y}\,{\cal R}(\theta)\,\ket{{\bf F}}{} \lab{rot1}\\
	&=&\!\!\!\!\sum_{n,m;q_x',q_y'}\!\!\!\!\braket1{q_x,q_y}{n,m}{\!P}\,
		\bra{P\!}{n,m}{\cal R}(\theta)\ket{n,m}{\!P} \braket{P\!}{n,m}{q_x',q_y'}1\,
		\braket1{q_x',q_y'}{{\bf F}}\nonumber\\
	&=&\!\!\!\!\sum_{n,m;q_x',q_y'}\!\!\!\!\braket1{q_x,q_y}{n,m}{\!P}\,e^{-\ii m\theta}
		\braket{P\!}{n,m}{q_x',q_y'}1\braket1{q_x',q_y'}{{\bf F}}{}{\qquad}\nonumber\\
	&=&\sum_{q_x',q_y'}R(q_x,q_y;q_x',q_y';\theta)\,F_{q'_x,q'_y} \lab{rot2}
\eea
with $n|_0^{2j}$, $m|_{-2j}^{2j}$ and $q'_x,q'_y|_{-j}^j$,
where the kernel of rotation is given by
\be 
	R(q_x,q_y;q_x',q_y';\theta):=\sum_{n,m}
		\Lambda^j_{n,m}(q_x,q_y)\,e^{-\ii m\theta}\,
		\Lambda^j_{n,m}(q_x',q_y')^*.  \lab{Rkernel1}
\ee
An alternative expression for this kernel can be written as \cite{APVW-II}
\be 
	\begin{array}{l} \displaystyle
		R(q_x,q_y;q_x',q_y';\theta):=\sum_{\mu,\mu'}
		\Psi_{n_x,n_y}(q_x,q_y)\, d^{n/2}_{\mu,\mu'}(2\theta)\, 
		\Psi_{n_x',n_y'}(q_x',q_y')^*, \\
		\mu=\onehalf(n_x-n_y),\ \mu'=\onehalf(n_x'-n_y');\quad 
				n_x+n_y=n=n_x'+n_y'. \end{array}		\lab{Rkernel2}
\ee	

When the input image is a matrix of real numbers, after rotation it
will remain a real matrix, because the rotation kernel 
\rf{Rkernel1} is real. Due to the orthonormality of the bases,
from \rf{Rkernel1} or \rf{Rkernel2}, we see that $R(q_x,q_y;q_{x'},q_{y'};0)
=\delta_{q_x,q_{x'}}\delta_{q_y,q_{y'}}$ is the unit transformation. 
Similarly, the product of two rotations satisfies group composition
$\sum_{q_{x'},q_{y'}} R(q_x,q_y;q_{x'},q_{y'};\theta_1)\*
R(q_{x'},q_{y'};q_{x''},q_{y''};\theta_2)
=R(q_x,q_y;q_{x''},q_{y''};\theta_1+\theta_2)$, and the inverse
is $R(q_x,q_y;q_{x'},q_{y'};-\theta)= R(q_{x'},q_{y'};q_x,q_y;\theta)^*$.
This shows that the matrices in the set \rf{Rkernel1} 
are unitary, ${\bf R}(-\theta)={\bf R}(\theta)^\dagger$, and is a
representation of the group \Lie{SO($2$)} of plane rotations; it is reducible in
the polar basis but irreducible in the Cartesian basis. As an example of
this rotation algorithm we provide Fig.\ \ref{fig:Monorot} that we proceed 
to examine and comment.


\subsection{The `Gibbs-{\it like\/}' oscillation phenomenon}

\begin{figure}[t]
\centering  
\centerline{\includegraphics[width=1.0\columnwidth]{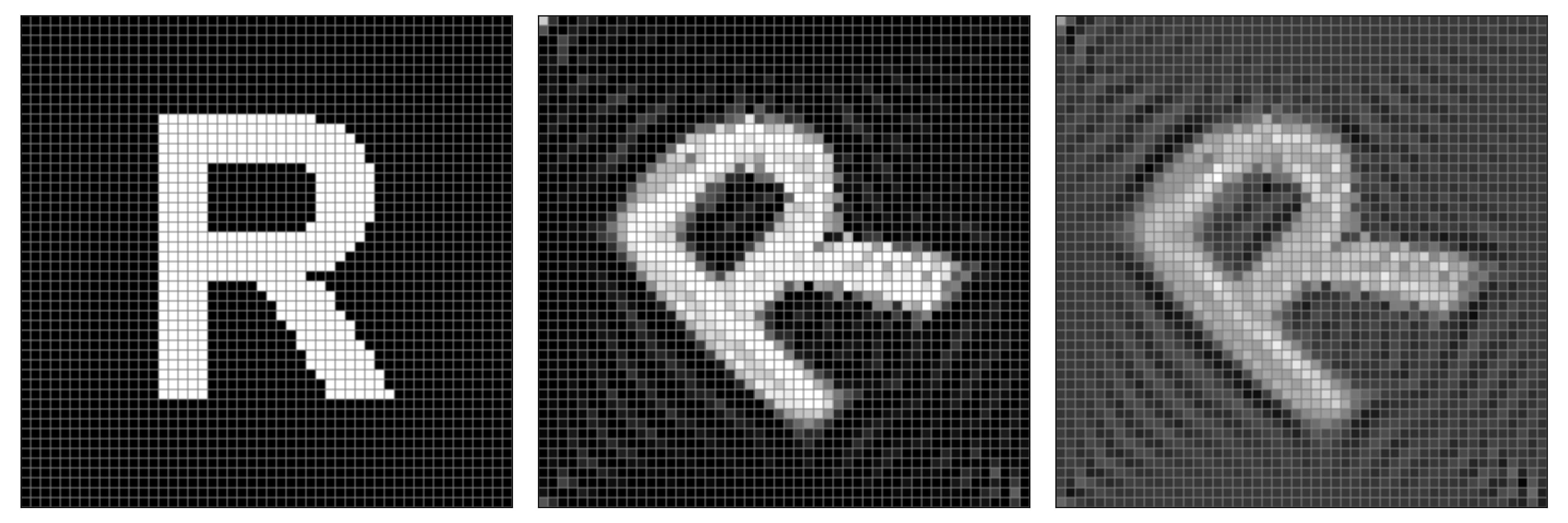}}
\caption[]{{\it Left\/}: A  $50\times 50$ monochromatic image (letter ``{\sf R}'') of 
pixel values 1, over a background of pixel values 0. 
{\it Center\/}: Rotation of 
this image by $\frac14\pi$ ($45^\circ$) with {\it clipping\/} of the over-
and under-shoots beyond the obligate range $[0,1]$. {\it Right\/}: Same rotation with
{\it normalization\/} of the pixel values to fit this range. Both output
images exhibit `Gibbs-{\it like\/}' oscillations due to the sharp
boundaries of the input image. Whereas clipping leads to a more contrasted
image, it also entails loss of information in pixel values; normalization 
retains information subject to translation and scale, although it appears 
less contrasted.}
\label{fig:Monorot}
\end{figure}

As is evident in Fig.\ \ref{fig:Monorot}, a 
ubiquitous consequence of the unitarity of the rotation
algorithm is that rotated images display {\it oscillations\/} 
due to `discrete discontinuities', i.e., large differences 
between values of neighboring pixels.
This oscillation pattern reminds us of that appearing
in another context: the summation of truncated Fourier series, 
known as the Gibbs phenomenon. Because their origin is different
(i.e., it is not a {\it type\/} of Gibbs oscillation),
we shall call these oscillations Gibbs-{\it like\/};
here we address this phenomenon for the first time.

Linear normalization is the main tactic to keep the grey-tone scale within 
the specified range $[0,1]$, between absence and saturation of the 
pixel values. We thus distinguish between the {\it data-image\/} 
whose pixel values can have any range, and the {\it screen-image\/} 
which is output for display, where pixel values are all within $[0,1]$. 
This tactic maps the data- to screen-values, ${\bf F}\mapsto\overline{\bf F}$, 
using the extreme values,
\be 
    F_{q_x,q_y} \mapsto \overline{F}_{q_x,q_y}=
		\frac{F_{q_x,q_y}-s_{N}}{S_{N}-s_{N}}, \quad
	\begin{array}{l} 
	s_{N}\,:=\min_{q_x,q_y}\,\{F_{q_x,q_y}\}<0,\\
	S_{N}:=\max_{q_x,q_y}\,\{F_{q_x,q_y}\}>1. 
	\end{array} \lab{minimax}
\ee
The screen-image will thus have a minimum of complete blacks 
and whites, with no information loss.

\begin{figure}[htbp]
\centering  
\centerline{\includegraphics[height = 0.8\textheight]{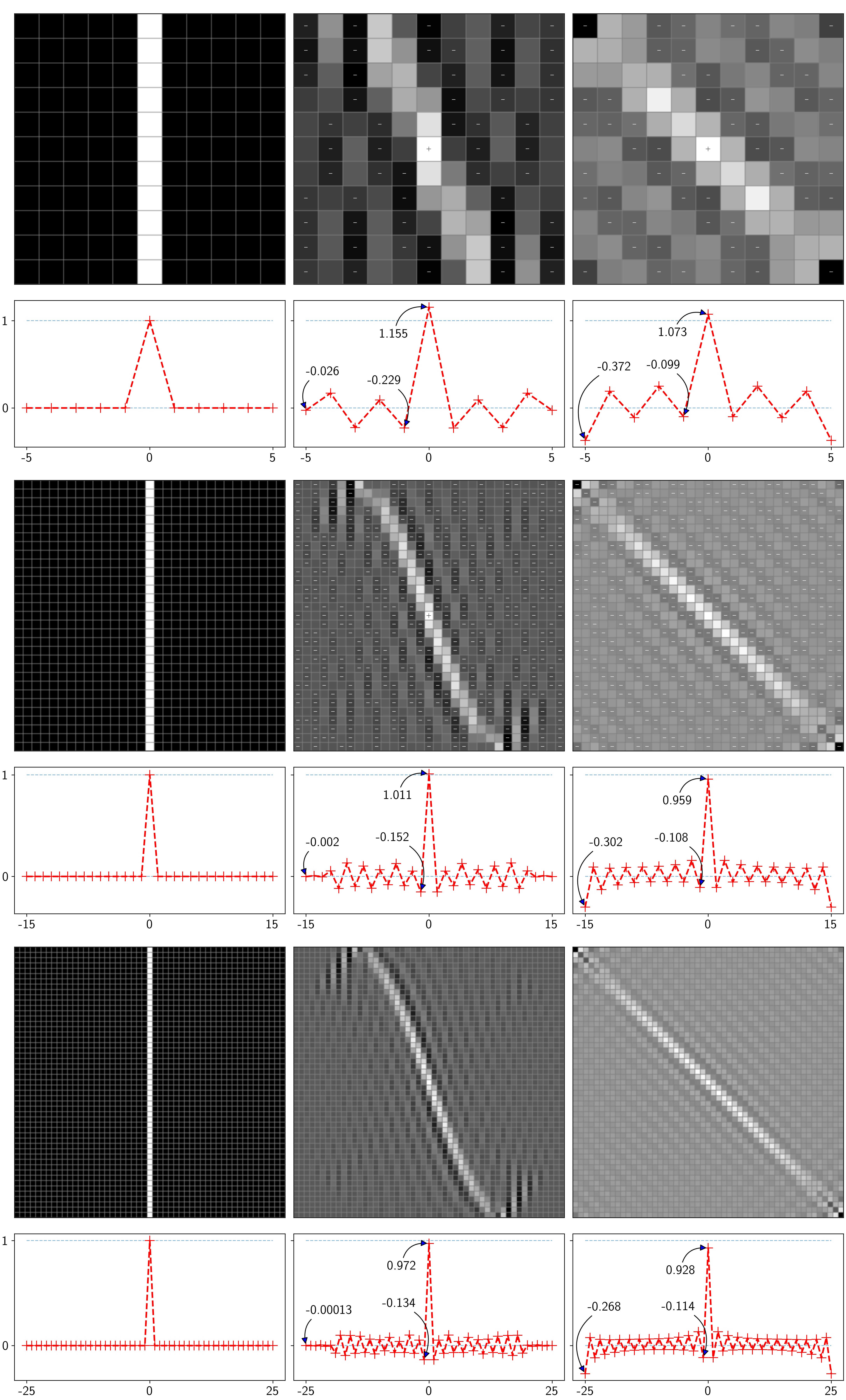}}
\caption[]{Unitary rotations of a centered Kronecker delta \rf{delta} 
for $N=11$, 31, 51, with their normalized rotated images.  
{\it Left\/}: the original discrete Kronecker function 
$\Delta^{\ssty(N)}_{\oof{x}}(q_x,q_y)$ of 0 and 1. {\it Center\/} and 
{\it Right\/}: their images rotated by  $\theta=\frac18\pi$ and 
$\frac14\pi$. Below each, the (non-normalized) 
values of  anti-diagonal pixels, with their largest 
over- and undershoots. For $N=11$ and 31, the  over- 
or under-shoots are marked by $+$ and $-$.}
\label{fig:Gibbs-0}
\end{figure}

\begin{figure}[htbp]
\centering  
\centerline{\includegraphics[height = 0.8\textheight]{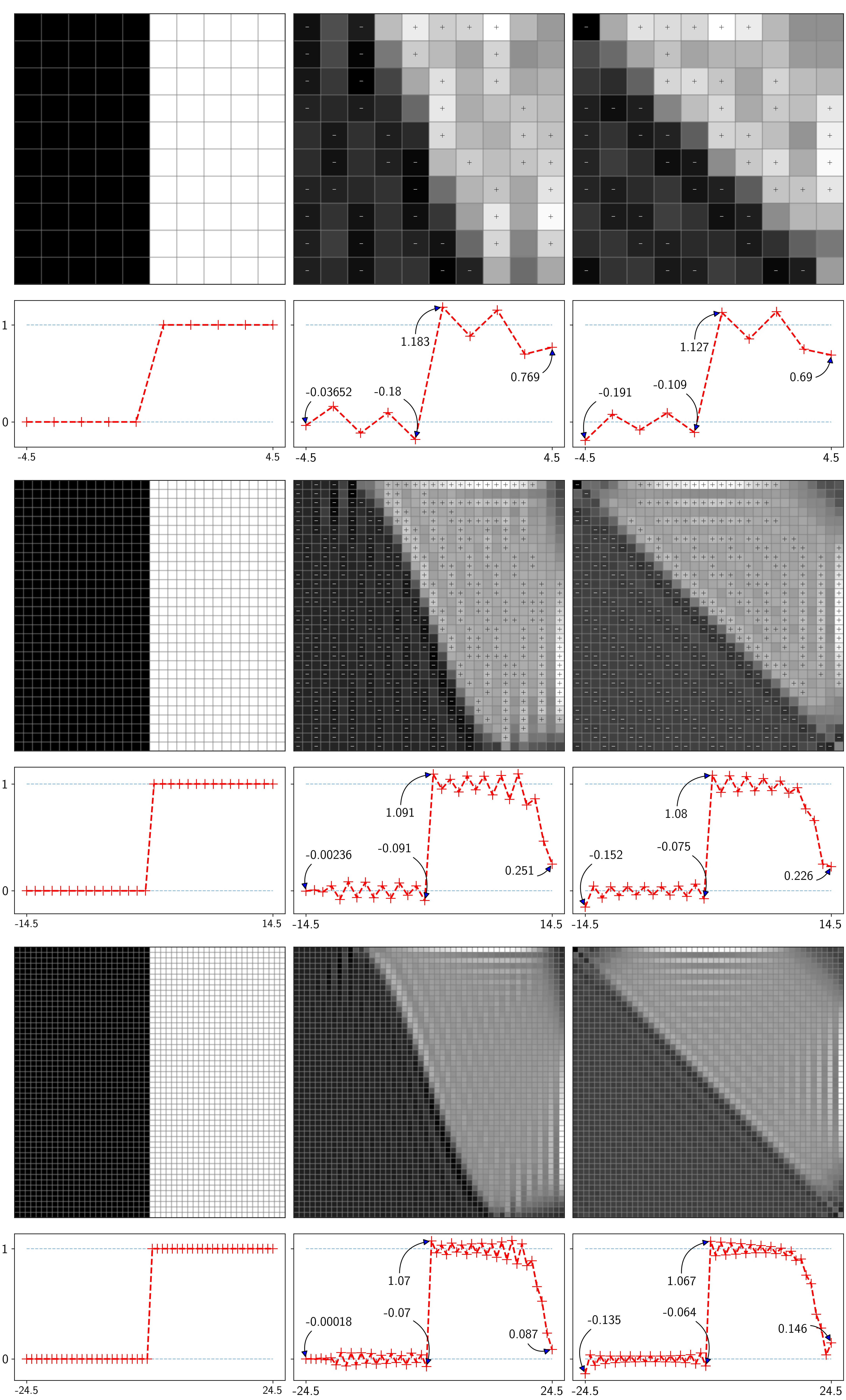}}
\caption[]{Unitary rotations of the unit step function \rf{step} 
for $N=10$, 30, 50, with normalized rotated images, organized as 
in Fig.\ \ref{fig:Gibbs-0}. {\it Left\/}: the step function 
$\Theta^{\ssty(N)}_{\oof{x}}(q_x,q_y)$. {\it Center\/} and {\it Right\/}:
rotatated images for $\theta=\frac18\pi$ and $\frac14\pi$.}
\label{fig:Gibbs-1}
\end{figure}

Whereas in the familiar  trigonometric Gibbs phenomenon, 
a unit step function $\Theta(q):=\onehalf(1+\hbox{sign}\,q)$ entails
a maximal overshoot of ${\sim}8.95\%$ at a positions ever closer 
to the discontinuity as the number of series terms grows 
(see \cite[Fig.\ 4.11 at p.\ 167]{Int-Transf}), to 
calibrate the Gibbs-{\it like\/} phenomenon for
$N\times N$ images under unitary rotation, let us consider a
Kronecker delta  at the center of the screen with an odd 
number $N=2j+1$ of pixel rows and columns,
\be 
	\Delta^{\ssty(N)}_{\oof{x}}(q_x,q_y):=\{0\hbox{ for } 
	q_y<0,\  1 \hbox{ at } q_y=0,\ 0\hbox{ for } q_y>0\},
 		\lab{delta}
\ee
and a  discrete step function along the $x$-axis for even 
$N=2j+1$ rows and columns
\be 
	\Theta^{\ssty(N)}_{\oof{x}}(q_x,q_y):=\{0\hbox{ for } 
		q_y\le-\onehalf,\ 1\hbox{ for }q_y\ge\onehalf\},
 			\lab{step}
\ee
shown for rotations of $\frac18\pi$ and $\frac14\pi$ in
Figs.\ \ref{fig:Gibbs-0} and \ref{fig:Gibbs-1} respectively.
The resulting oscillations are recognizably parallel to the
lines of discontinuity. To evince this oscillation behavior 
we plot below the pixel values along the main 
{\it anti\/}-diagonal of each screen 

A first observation is that, unlike the continuous case, the
over- and under-oscillations are not symmetrical. This is
because while in the Gibbs case $e^{0\ii x}=1$ belongs to the Fourier 
series terms and allows the addition of constants, in the discrete
case $\Psi_{0,0}(q_x,q_y)$ is {\it not\/} a constant that could be freely
added to all pixel values. 
Most importantly, we note that the over- and under-shoots {\it diminish\/} 
with increasing $N$: as the pixellation becomes finer, the finite oscillator 
states approximate those of continuous systems ---as expected. 
Yet, although the delta and step functions lead to rotated images that 
are relatively simple to write analytically, they are not very useful
to elucidate certain systematic features of their rotated images on the
pixellated screen, such as the the behavior on and near the screen vertices
and edges, as can be seen in those figures.


\section{Three-color pixellated images}   \label{sec:three}

With a qualitative understanding of oscillations due to discrete
Kronecker deltas and step functions, we now extend the algorithm 
from rotations of monochromatic finite 
$N\times N$ pixellated images presented in the previous 
section, to polychromatic images within the standard three-color 
scheme of RGB (Red-Green-Blue) technology. Each channel, R, G or B, 
is a monochromatic image; polychromatic images (indicated by 
sans serif font) are the direct sum of the three components 
where the pixel values for each color are all in the range $[0,1]$,
\be
	\FFF={\bf F}^\ssr{R}\oplus{\bf F}^\ssr{G}\oplus{\bf F}^\ssr{B},
	\quad  \braket{1\!}{q_x,q_y}{{\FFF}}{} =  \left[\begin{array}{l}
		\scriptstyle\braket{1\!}{q_x,q_y}{{\bf F^\ssr{R}}}{} \\[-3pt]
	    \scriptstyle\braket{1\!}{q_x,q_y}{{\bf F^\ssr{G}}}{} \\[-3pt] 
	    \scriptstyle\braket{1\!}{q_x,q_y}{{\bf F^\ssr{B}}}{}\end{array}\right]\!\!,
	    \quad  0\le \braket{1\!}{q_x,q_y}{{\bf F^\ssr{X}}}{} \le1,
	    		\lab{tres-colores} 
\ee
where X stands for R, G or B, with values 0 for absence and 1 for saturation 
in each color.

Under rotation we let each component undergo the same transformation 
\rf{rot1}--\rf{Rkernel1} detailed in the previous section,
\be 
	{\FFF}^{(\theta)}:=
	{\cal R}(\theta){\FFF}= {\cal R}(\theta){\bf F}^\ssr{R}\oplus
		{\cal R}(\theta){\bf F}^\ssr{G}\oplus
		{\cal R}(\theta){\bf F}^\ssr{B}.   \lab{rota-3c}
\ee 

The RGB color scheme is an industrial standard \cite{4-AU} that entails 
an additive colorimetric scale, that is, each of the elements of each channel 
sums to define an individual color \cite{1-AU, 2-AU}. Physically, each of 
the colors in the RGB scheme is associated with the way the human eye 
perceives and processes light with the photoreceptor cells in the retina 
\cite{3-AU}. The chromatic response is intrinsic to the aggregation of 
the three channels and independent of the luminescence, which means that 
the white point of the scale is the algebraic sum of the three channel 
values, and independent of the illumination \cite{12-AU}. This color 
scheme is strictly positive within a range of bits or bytes displayed 
in digital or analogue devices \cite{5-AU, 6-AU}. 
Although pixel values outside the standard range have been also used as
representing complementary colors \cite{7-AU, 8-AU}, we forego this
alternative interpretation because it would introduce dubious changes
in the color rendering of black or white backgrounds. 

\begin{figure}[htbp]
\centering
\centerline{\includegraphics[width=\columnwidth]{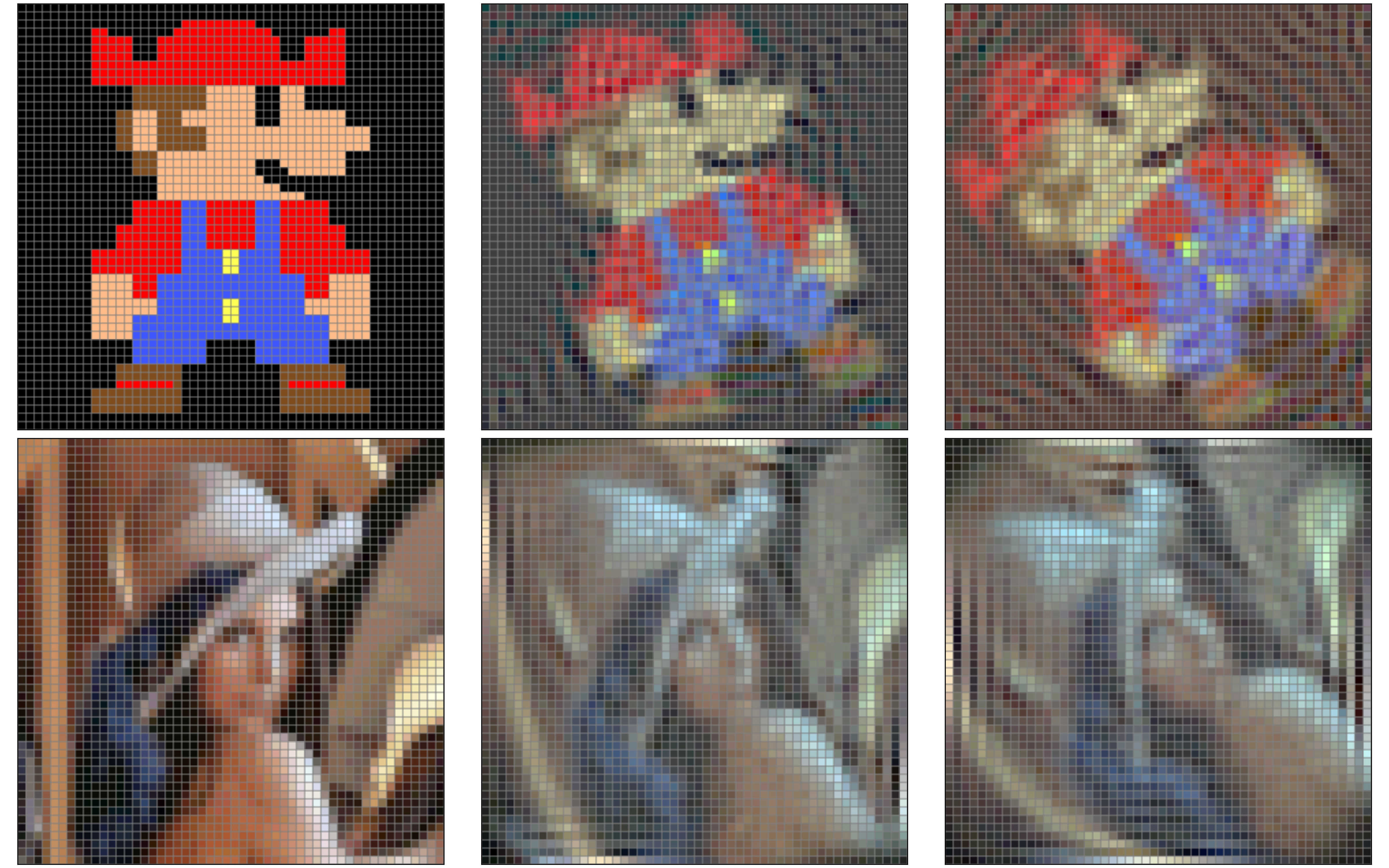}}
\caption[]{{\it Left\/}: Two RGB $52\times52$ pixellated images
(above a {\it pseudo\/}-Mario, below, the well-known {\it Lena}).
{\it Center\/} and {\it Right\/}: Their unitary rotation by
$\theta = \frac18\pi$ and $\frac14\pi$ with the normalization 
\rf{minimax1}; the data image of the former can be further 
rotated by $\frac18\pi$ to obtain the data-image of the latter, 
because the unitary transformations satisfy composition.} 
\label{fig:rota-fin}
\end{figure}

Here we address the {\it joint\/} display of the three data-image
matrices to acceptably transform color images. The normalization 
of pixel values \rf{minimax} now requires the extreme values among 
the {\it three\/} colors to define, for $Y\in\{R,G,B\}$, 
\be
	 F_{q_x,q_y}^{\ssr Y} \mapsto \overline{F}_{q_x,q_y}^{\ssr Y}
		=\frac{F_{q_x,q_y}^{\ssr Y}-s_{N}}{S_{N}-s_{N}},\quad
	\begin{array}{l}
	s_{N}\,:=\min_{X;\,q_x,q_y}\,\{F_{q_x,q_y}^{\ssr X}\}<0,\\
	S_{N}:=\max_{X;\,q_x,q_y}\,\{F_{q_x,q_y}^{\ssr X}\}>1. 
		\end{array} \lab{minimax1}
\ee
Two examples of this unitary rotation algorithm for 
three-color images are presented in Figs.\ \ref{fig:rota-fin},
for images with high and low pixel discontinuity contrasts. 

We should remark that the computation of all figure rotations
in this report has been performed with an algorithm
that enhances the speed of the previously used rotation 
algorithms \cite{APVW-II,rectangular} that needs the computation 
of the special functions in the kernel \rf{Rkernel1}. Previous 
reports of unitary transformations did not focus on the computational 
efficiency because one was more interested in the mathematical 
features of the transformations. Yet for example in \cite{rectangular} 
a grayscale rectangular image of $61\times 37 = 2257$ pixels was calculated 
in $\sim$78 hours, here each of the three layers of Fig.\ 
\ref{fig:rota-fin} has $52\times 52 = 2704$ pixels that were
computed in  $\sim$250 seconds, while the full RGB 
image took $\sim$12 minutes to complete. This
improvement is due to two factors: first, the optimization of 
the routine to calculate the elements in \rf{Rkernel1} was done 
through a three-term recurrence relation (instead of library 
functions), and that a high-level scripting language such as Julia
\cite{10-AU} was used, allowing the implementation of a 
syntactically natural transformation algorithm that simultaneously 
profits from the multi-threading in modern CPU's. For more technical 
details, such as the binomial overflow as $N$ grows, the algorithm 
we used in all the previous figures is provided by an {\it Open Source\/} 
code shared on a public repository \cite{11-AU}, to fulfill the 
principle of reproducibility for the results here presented.

\section{Conclusions}           \label{sec:four} 

We remarked above that the unitary rotation algorithm for pixellated images 
based on the finite oscillator model is unique, and can be inverted or 
concatenated.  Yet we should underline that the unitary rotation algorithm 
requires that the value at every pixel in the rotated image depend on the 
values of {\it all\/} pixels of the input image. We are thus dealing 
with an $\mathcal{O}(N^{4})$ dependence in the number of operations 
needed to compute a rotation. 
The obvious advantage of information conservation comes thus at the 
price of a lengthier computation time. For commercial 
applications, where $N\approx 10^3$ is common, the more common
image rotation algorithms based on interpolation in smaller
regions of an image, have none of the former advantages, nor the 
last disadvantage.

The results presented for three basic colors of course also apply
to any number of quantities that may be attached to the pixel
values in pre-ordained ranges. The unitary rotations can also be made 
to act on the latter quantities {\it covariantly}, such as would be the case
for a discrete and finite {\it vector field}, whose two components will
rotate together with the data-image.  Also, three- and higher-dimensional 
pixellated (`voxellated') images, subject to rotations \cite{D-dim-rot}
can be subject to considerations similar to those analyzed here.

Lastly, we should place the group of rotations of pixellated
screens within the wider context of the finite oscillator model,
which uses basically the rotation Lie subalgebra chain $\Lie{so($3$)}
\supset\Lie{so($2$)}$ to provide the pixel coordinates and 
a phase space interpretation for the system, as
opposed to their usual numbering through the $N$-point Fourier
transform. In the latter, the position space forms a discrete torus
whose two discrete $(x,y)$ coordinates curl into circles, but 
where rotations between them cannot be applied without breaking
the torus. In the finite oscillator model, the 
manifold harboring the coordinates is a sphere, which can be
freely rotated and also gyrated in phase space, among other
linear canonical transformations.

\section*{Acknowledgments}

We thank the support of the Universidad Nacional Aut\'onoma
de M\'exico through the {\sc dgapa-papiit} project AG100120
{\it \'Optica Matem\'atica}. A.R.\ Urz\'ua acknowledges the 
support of the National Council for Science and Technology
({\sc conacyt}) through the {\it Becas Postdoctorales\/} program (2021).



\begin{thebibliography}{99}

\bibitem{AW} N.M.\ Atakishiyev, L.E.\ Vicent and K.B. Wolf,
	 Continuous vs.\ discrete fractional Fourier transforms, 
	 \jour{J.\ Comp.\ Appl.\ Math.}{107}{73--95}{1999}.

\bibitem{APVW-II} N.M.\ Atakishiyev, G.S.\ Pogosyan, 
	L.E.\ Vicent, and K.B.\ Wolf,
	Finite two-dimensional oscillator II.\ The radial model,
	\jour{J.\ Phys.\ A}{34}{9399--9415}{2001}.
	
\bibitem{VW-rot} K.B.\ Wolf and L.E.\ Vicent,
	The Fourier U(2) group and separation of discrete variables,	
	\jour{Sigma}{7}{053, 18p.}{2011}.
	
\bibitem{APVW-I} N.M.\ Atakishiyev, G.S.\ Pogosyan, L.E.\ Vicent 
	and K.B.\ Wolf, Finite two-dimensional oscillator. I: The Cartesian
	 model, \jour{J.\ Phys.\ A}{34}{9381--9398}{2001}.			
	
\bibitem{rectangular} A.R.\ Urz\'ua and K.B.\ Wolf,
	Unitary rotation and gyration of pixellated images on rectangular screens,
	\jour{J.\ Opt.\ Soc.\ Am.\ A}{33}{642--647}{2016}.
	
\bibitem{finite-model} N.M.\ Atakishiyev, G.S.\ Pogosyan, and K.B.\ Wolf,
	Finite models of the oscillator,
	\jour{Phys.\ Part.\ Nuclei}{36}{521--555}{2005}.
	
\bibitem{Bied-Louck} L.C.\ Biedenharn and J.D.\ Louck, Angular Momentum in
	Quantum Physics.\ Theory and Application. Encyclopedia of Mathematics
	and its Applications, Vol.\ 8 (Addison-Wesley Publ.\ Co., Reading, Mass.,1981).

\bibitem{11-AU} A.R.\ Urz\'ua (2022), Polychromatic Unitary Rotations, 
	version 0.1.0 [Computer software], in https://github.com/rurz/UnitRots.
	
\bibitem{Int-Transf} K.B.\ Wolf, Integral Transforms in Science and
	Engineering (Plenum Press, New York, 1979).
 
\bibitem{color-constancy} C.\ van Trigt, 
	Open problems in color constancy: discussion,
	\jour{J.\ Opt.\ Soc.\ Am.\ A}{31}{338--347}{2014}
	
\bibitem{1-AU} R.G.\ Kuehni, Color space and its divisions, 
	\jour{Color Research \& Application}{26}{209--222}{2001}.

\bibitem{2-AU} I.\ Lissner and P.\ Urban, Toward a unified color space for
	perception-based image processing, \jour{IEEE Transactions on Image
	Processing}{21}{1153--1168}{2012}.

\bibitem{3-AU} G.\ Wyszecki and W.S.\ Stiles, {\it Color Science\/} Vol.\ 8
	(Wiley, New York, 1982).

\bibitem{4-AU} S.\ S\"usstrunk, R.\ Buckley, and S.\ Swen, Standard rgb 
	color spaces, {\it Color and Imaging Conference 1999}, {127--134} (1999).

\bibitem{5-AU} H-R.\ Kang,{\it Color Technology for Electronic Imaging Devices},
	(SPIE Optical Engineering Press, Bellingham, Wash., 1997).

\bibitem{6-AU} Z.\ Shen. Three-Color Tunable Organic Light-Emitting Devices,
	\jour{Science}{276}{2009--2011}{1997}.
	
\bibitem{8-AU} R.W.\ Pridmore, Complementary colors: The structure of wavelength 
	discrimination, uniform hue, spectral sensitivity, saturation, chromatic 
	adaptation, and chromatic induction,
	\jour{Color Research \& Application}{34}{233--252}{2009}.

\bibitem{7-AU} R.W.\ Pridmore, Complementary colors: A literature review,
	\jour{Color Research \& Application}{46}{482--488}{2020}.
	
\bibitem{9-AU} A.R.\ Urz\'ua, ``Rotaciones y giraciones en pantallas cartesianas 
	rectangulares" (M.Sc.\ thesis, UNAM, 2016), https://repositorio.unam.mx/contenidos/95764.

\bibitem{10-AU} J.\ Bezanson, A.\ Edelman, S.\ Karpinski, and V.B.\ Shah,  
	Julia: A Fresh Approach to Numerical Computing. In:
	\jour{SIAM Review}{59}{65--98}{2017}.  https://doi.org/10.1137/141000671.

	
\bibitem{12-AU} H.\ Grassmann, 
	Zur Theorie der Farbenmischung, \jour{Ann.\ Phys.\ Chem.}{165}{69}{1853}. 

\bibitem{D-dim-rot} G.\ Krötzsch, K.\ Uriostegui and K.B.\ Wolf, Unitary 
	rotations in two-, three-, and $D$-dimensional Cartesian data arrays, 
	\jour{J.\ Opt.\ Soc.\ Am.\ A}{31}{1531--1535}{2014}.

\end{thebibliography}
\end{document}